\documentclass[epj]{webofc}
\usepackage[varg]{txfonts}   
\woctitle{XLV International Symposium on Multiparticle Dynamics}
\begin{document}
\title{LHCb results from proton ion collisions}
%
%

\author{Laure Massacrier on behalf of the LHCb Collaboration\inst{1,2}\fnsep\thanks{\email{massacri@lal.in2p3.fr}}      
}

\institute{LAL, Univ. Paris-Sud, CNRS/IN2P3, Universit\'e Paris-Saclay, Orsay, France
\and
    IPNO, Univ. Paris-Sud, CNRS/IN2P3, Universit\'e Paris-Saclay, Orsay, France
          }

\abstract{%
  Proton-lead and lead-proton data taking during 2013 has allowed LHCb to expand its physics program to heavy ion physics. Results include the first forward measurement of Z production in proton-lead collisions as well as a measurement of the nuclear modification factor and forward-backward production of prompt and displaced J/$\psi$, $\psi$(2S) and $\Upsilon$. Angular particle correlations have also been measured for events of varying charged particle activity.
}
\maketitle
\section{Introduction}
\label{intro}

Ultra-relativistic heavy-ion collisions are used to study the nuclear matter at high temperature and pressure where the formation of the Quark Gluon Plasma (QGP), a state of matter which consists of asymptotically free quarks and gluons, occurs. Open heavy flavours and quarkonia are produced at the early stages of the collision and might interact with the deconfined medium, making them ideal probes of the QGP. It was indeed predicted that in hot nuclear matter, charmonia are suppressed due to color screening of the heavy quarks potential \cite{Matsui:1986dk}. Quarkonia can also be suppressed in absence of the QGP formation by Cold Nuclear Matter (CNM) effects. Proton-Nucleus collisions (pA or Ap), which are interesting by themselves, are therefore essential to interpret Nucleus-Nucleus data in order to disentangle QGP effects from CNM effects in those collisions. The main CNM effects affecting quarkonium production include 
initial-state nuclear effects on the parton densities (shadowing) \cite{Albacete:2013ei}, the initial-state parton energy loss and final-state energy loss (coherent energy loss) \cite{Arleo:2012rs}, the final-state absorption of the pre-resonant heavy quark pair by the spectator nucleons (nuclear absorption) albeit small at LHC \cite{Ferreiro:2013pua} and the final-state interaction of the quarkonium with the produced medium (comovers) \cite{Ferreiro:2014bia}. \\
Proton-nucleus collisions are also useful for the determination of nuclear parton distribution functions (nPDF) \cite{Salgado:2011wc}. The measurement of the Z electroweak boson in pA and Ap collisions in the LHCb acceptance permits to probe low $x_{A}$ values\footnote{$x_{A}$ is the momentum fraction of a certain parton inside a given nucleon bound in the nucleus A.}(2 $\times$ $10^{-4}$ - 3 $\times 10^{-3}$) in the forward case, and high $x_{A}$ value (0.2-1.0) in the backward case for an energy scale $Q^{2}$ = $M_{Z}^{2}$. Such a measurement (the first performed at the LHC in pA collisions) should offer large constraining power to nPDF fits especially at small $x_{A}$.
Finally, two-particle angular correlations are interesting tools to probe collective effects in the dense environment of high-energy collisions. Thanks to the high particle density and multiplicities reached in pA collisions, of similar size to that of non central nucleus-nucleus collisions (AA), LHCb can investigate at forward rapidity the long-range correlation on the near side (ie the so-called ridge \cite{Khachatryan:2010gv}) which was already observed in proton-proton (pp), proton-lead (pPb) and lead-lead (PbPb) collisions at mid-rapidity.

\section{The LHCb detector and data taking}
\label{sec-1}
The LHCb detector is a single arm spectrometer in the forward region. It is fully instrumented in its angular acceptance and covers the pseudo-rapidity range 2 $< \eta <$ 5. Initially designed for b-physics, the LHCb detector is becoming a general purpose detector by carrying out an extensive heavy-ion program (including pA, AA data taking in the collider mode and a rich fixed target program). More details on the LHCb detector apparatus can be found in \cite{Alves:2008zz,Aaij:2014jba}. In 2013, LHCb collected pPb and Pbp data at a center-of-mass (CMS) energy of $\sqrt{s_{NN}}$ = 5 TeV. In such asymmetric collisions, the nucleon-nucleon center-of-mass system is shifted by 0.47 unit of rapidity in the direction of the proton beam. In the forward (backward) configuration pPb (Pbp), the proton (lead) beam traverses LHCb from the vertex locator to the muon system, respectively. The LHCb acceptance is 1.5 $< \rm{y}_{CMS} <$ 4.0 in the forward configuration and -5.0 $< \rm{y}_{CMS} <$ -2.5 in the backward configuration, leading to a common rapidy range coverage for both configurations of 2.5 $<\mid  \rm{y}_{CMS} \mid <$ 4.0. Data collected in the forward configuration amounts to $L_{int}$ = 1.1 nb$^{-1}$ while in the backward configuration it amounts to $L_{int}$ = 0.5 nb$^{-1}$. \\

\section{J/$\psi$, $\psi(2S)$, $\Upsilon$ production in pPb/Pbp and CNM studies}

\begin{figure*}
\centering
 \includegraphics[width=5.0cm,clip]{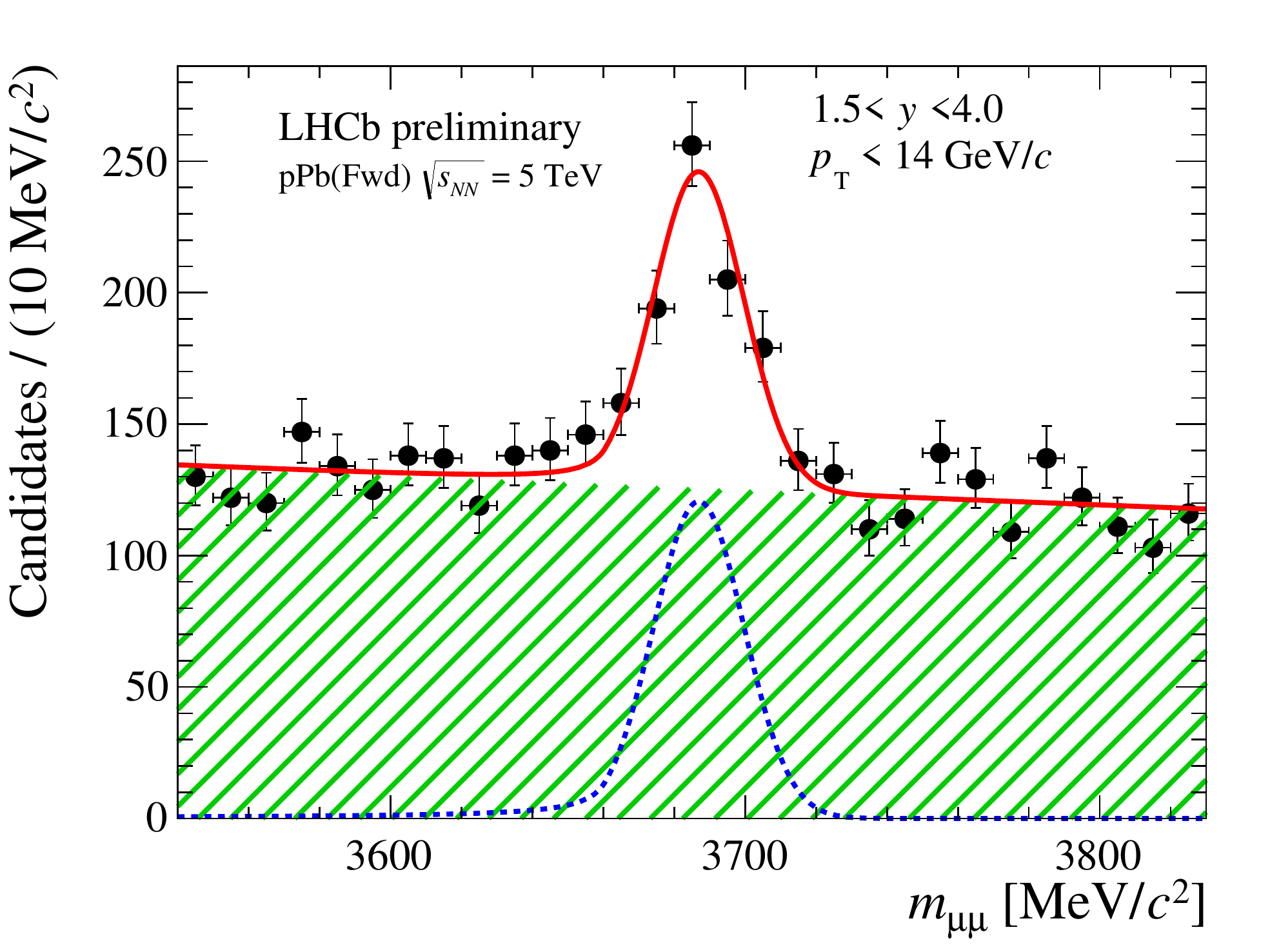}
 \includegraphics[width=5.0cm,clip]{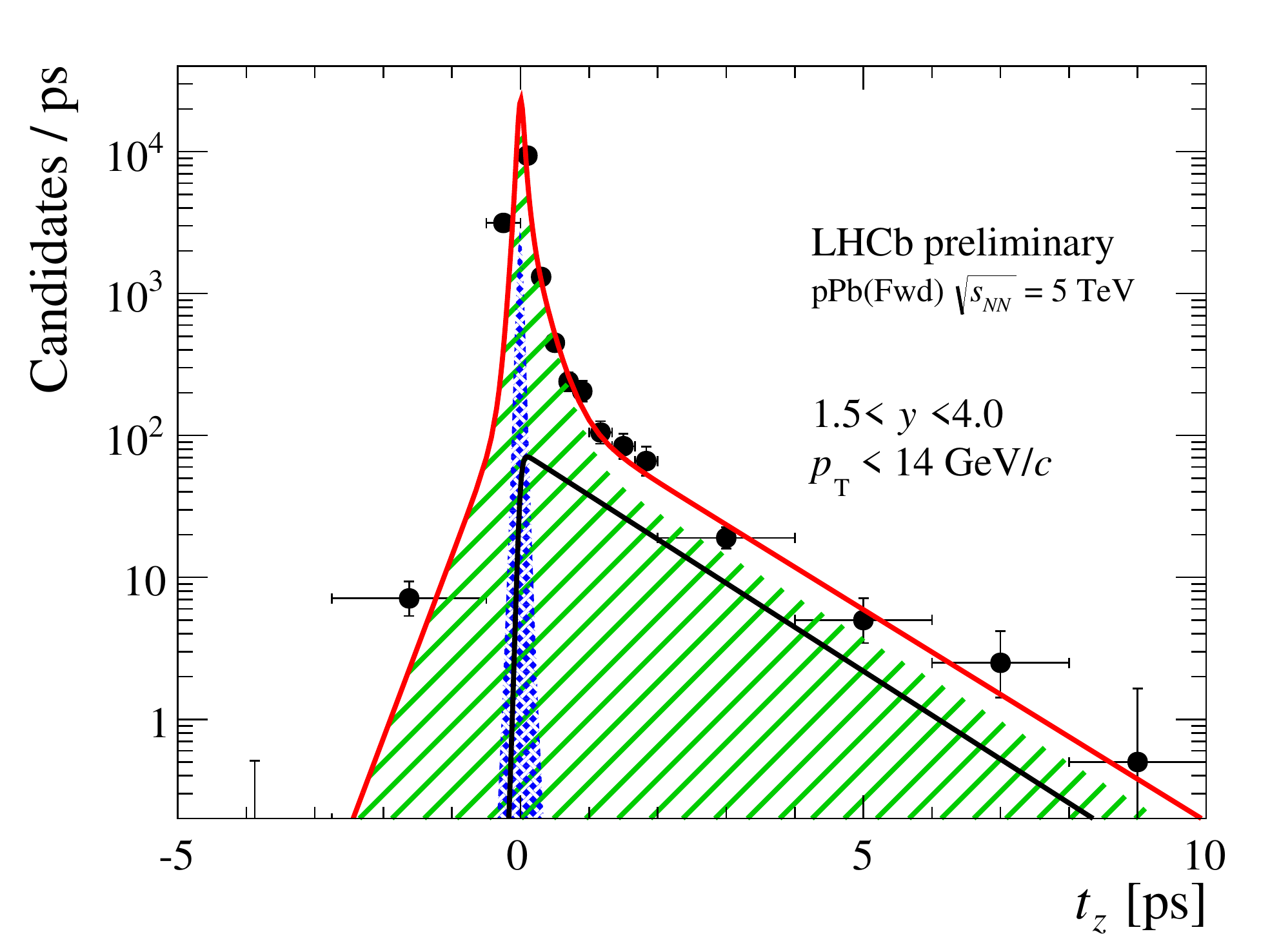}
\caption{Projections of the $\psi$(2S) fit results to the dimuon invariant mass (left) and pseudo-proper time $t_{z}$ (right) in pPb forward configuration (1.5 $< \rm{y}_{CMS} <$ 4.0). For the mass projection, the total fitted function is the red line, the signal distribution is the blue dotted line and the combinatorial background is represented by the green hatched area. For the $t_{z}$ projection, the same representation is adopted and in addition the $\psi$(2S) from b-hadron signal is given by the black line.}
\label{fig-1}       
\end{figure*}

J/$\psi$, $\psi$(2S) and $\Upsilon$ are studied in the dimuon final state with a dimuon transverse momentum restricted to $p_{T} < 14$ GeV/$c$ for charmonia and to $p_{T} < 15$ GeV/$c$ for bottomonia. Prompt J/$\psi$ and $\psi$(2S) can be disentangled from J/$\psi$ and $\psi$(2S) coming from b-hadron decays thanks to the excellent vertexing capability of LHCb. The yields of prompt charmonia and charmonia from b-hadron decays are obtained, in each kinematic bin, from a simultaneous fit of the dimuon invariant mass and pseudo-proper time distributions (see Fig. \ref{fig-1} for the $\psi$(2S) signal). An unbinned extended maximum likelihood fit to the dimuon invariant mass of the $\Upsilon$ candidates was performed to determine the signal yields of $\Upsilon$(1S), $\Upsilon$(2S) and $\Upsilon$(3S). While the three $\Upsilon$ resonances are observed in the forward configuration, only $\Upsilon$(1S) gives a significant signal in the backward configuration. Nuclear effects are usually quantified by the nuclear modification factor $R_{\rm{pA}}$ and the forward to backward ratio $R_{\rm{FB}}$. $R_{\rm{pA}}$ is defined as the production cross-section of a given particle in pA collisions divided by its production cross section in pp collisions at the same center-of-masse energy, and scaled by the atomic mass number A of the nuclei, while $R_{\rm{FB}}$ is the ratio of the production cross-section of a given particle in pA over its production cross-section in Ap configuration, measured in the same absolute center-of-mass rapidity range:
\begin{equation}
R_{\rm{pA}} = \frac{1}{A}\frac{\rm{d}^{2}\sigma_{\rm{pA}}(y,p_{T})/\rm{d}\sigma \rm{d}y}{\rm{d}^{2}\sigma_{\rm{pp}}(y,p_{T})/\rm{d}\sigma \rm{d}y}, \hspace{1 true cm} R_{\rm{FB}} = \frac{\sigma_{\rm{pA}}(+\mid y \mid, p_{\rm T})}{\sigma_{\rm{pA}}(-\mid y \mid, p_{\rm T})}.
\end{equation}
$R_{\rm{FB}}$ has the advantage to not rely on the pp reference cross-section and that part of the experimental systematic uncertainties and theoretical scale uncertainties cancel. \\
To determine the nuclear modification factor $R_{\rm{pPb}}$ of the J/$\psi$, $\psi$(2S) and $\Upsilon$(1S), their pp reference production cross-section at $\sqrt{s_{NN}}$ = 5 TeV are required. Since no data at this energy were available so far\footnote{A first data dating period in pp collisions at $\sqrt{s}$ = 5 TeV has just been completed end of November 2015 by the LHC.}, the reference cross-sections for J/$\psi$ and $\Upsilon$(1S) are obtained by a power-law fit to existing LHCb measurements at 2.76, 7 and 8 TeV\cite{ALICE:2013spa,LHCb:08082014sca}. To get the $\psi$(2S) reference pp cross-section at $\sqrt{s}$~=~5~TeV, the following assumption has been made:
\begin{equation}
\frac{\sigma_{\rm{pp}}^{J/\psi}(5 \hspace{0.1 true cm} \rm{TeV})}{\sigma_{\rm{pp}}^{\psi(2S)}(5 \hspace{0.1 true cm} \rm{TeV})} \approx \frac{\sigma_{\rm{pp}}^{J/\psi}(7 \hspace{0.1 true cm} \rm{TeV})}{\sigma_{\rm{pp}}^{\psi(2S)}(7 \hspace{0.1 true cm} \rm{TeV})},
\end{equation}
assuming that the systematic uncertainty on this hypothesis is negligible compared to the statistical uncertainty on the $\psi$(2S) measurement in pA collisions. The $\psi$(2S) and J/$\psi$ production cross-sections at $\sqrt{s}$~=~7~TeV are taken from the LHCb measurements in Refs \cite{Aaij:2011jh,Aaij:2012ag}. Under this assumption, the $\psi$(2S) nuclear modification factor can be derived from the J/$\psi$ nuclear modification factor\cite{Aaij:2013zxa}, with the following equation:
\begin{equation}
R_{\rm{pPb}}^{\psi(2S)} = \frac{\sigma_{\rm{pPb}}^{\psi(2S)}(5 \hspace{0.1 true cm} \rm{TeV})}{\sigma_{\rm{pPb}}^{J/\psi}(5 \hspace{0.1 true cm} \rm{TeV})} \frac{\sigma_{\rm{pp}}^{J/\psi}(5 \hspace{0.1 true cm} \rm{TeV})}{\sigma_{\rm{pp}}^{\psi(2S)}(5 \hspace{0.1 true cm} \rm{TeV})} \times R_{\rm{pPb}}^{J/\psi} \approx \frac{\sigma_{\rm{pPb}}^{\psi(2S)}(5 \hspace{0.1 true cm} \rm{TeV})}{\sigma_{\rm{pPb}}^{J/\psi}(5 \hspace{0.1 true cm} \rm{TeV})} \frac{\sigma_{\rm{pp}}^{J/\psi}(7 \hspace{0.1 true cm} \rm{TeV})}{\sigma_{\rm{pp}}^{\psi(2S)}(7 \hspace{0.1 true cm} \rm{TeV})} \times R_{\rm{pPb}}^{J/\psi}
\end{equation}

\begin{figure*}
\centering
 \includegraphics[width=5.0cm,clip]{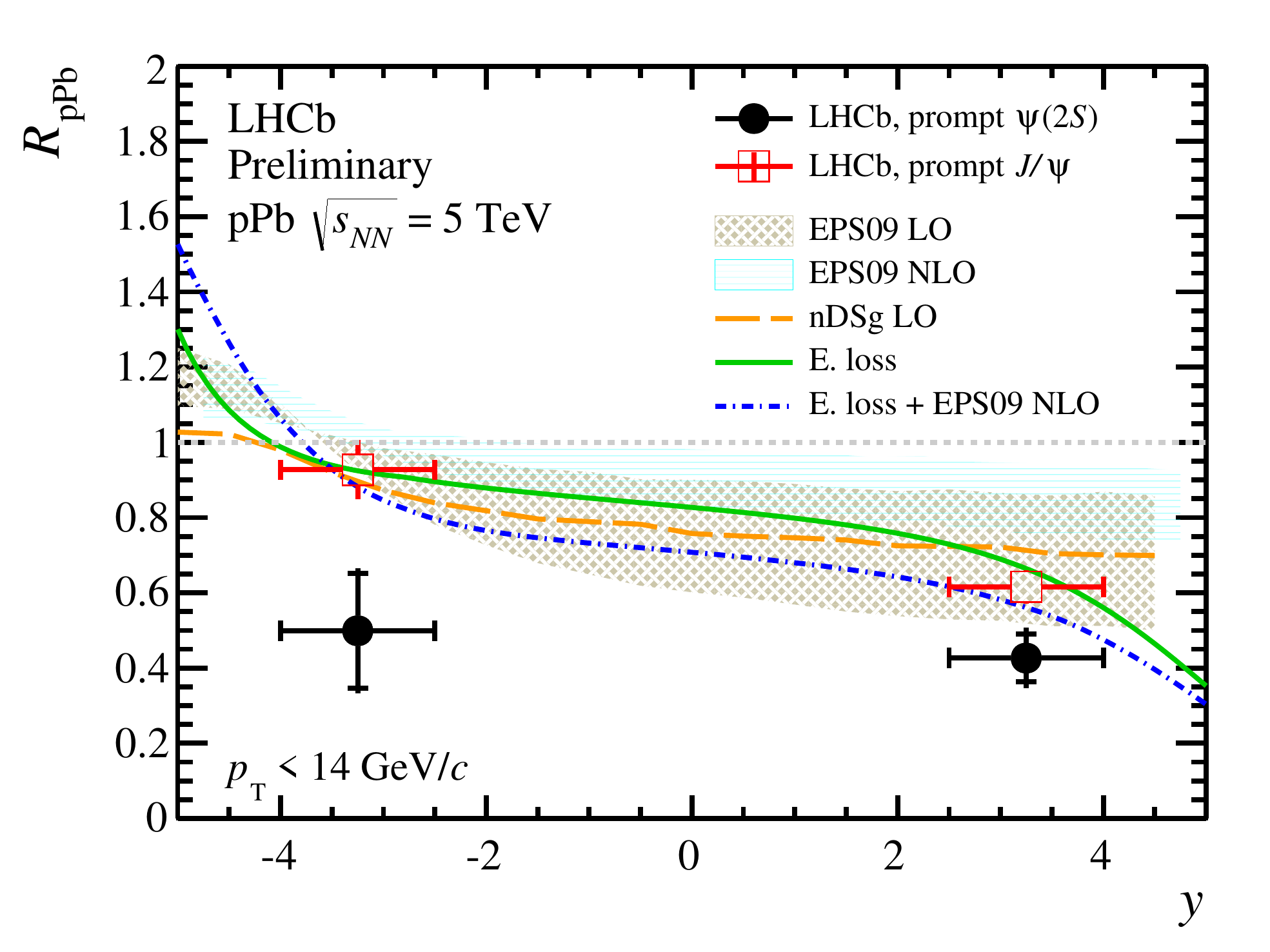}
 \includegraphics[width=5.0cm,clip]{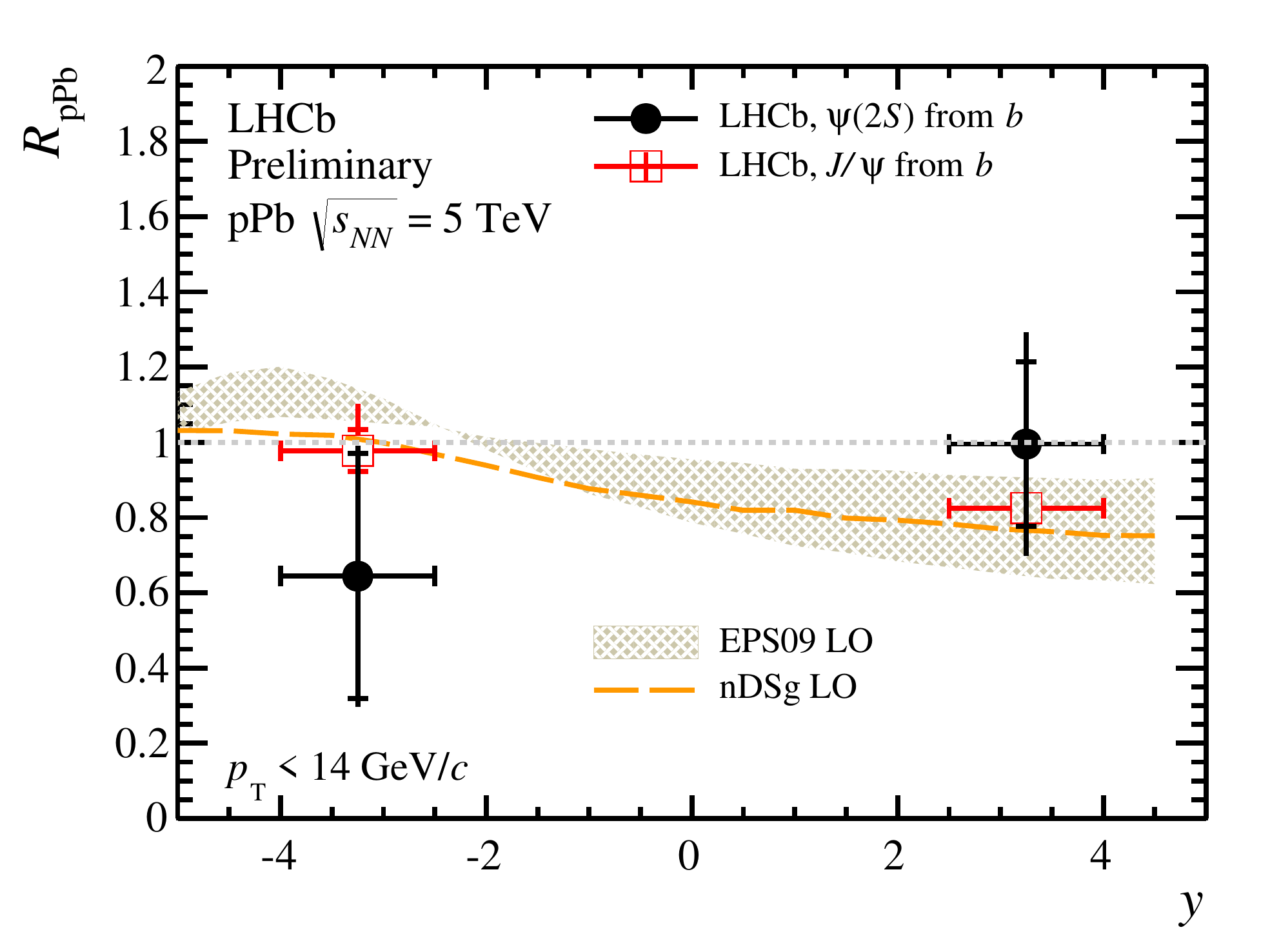}
\caption{Nuclear modification factor $R_{\rm{pPb}}$ as a function of rapidity for prompt J/$\psi$ and $\psi$(2S) (left) and for J/$\psi$ and $\psi$(2S) from b-hadrons (right). Results are compared with theoretical models from Refs. \cite{Ferreiro:2013pua,delValle:2014wha} (yellow dashed line and brown band), from Ref. \cite{Albacete:2013ei} (blue band) and from Ref. \cite{Arleo:2012rs} (green solid and blue dash-dotted lines). Only the models from \cite{Ferreiro:2013pua,delValle:2014wha} are available for $\psi$(2S) from b. } 
\label{fig-2}       
\end{figure*}

\begin{figure*}
\centering
 \includegraphics[width=5.0cm,clip]{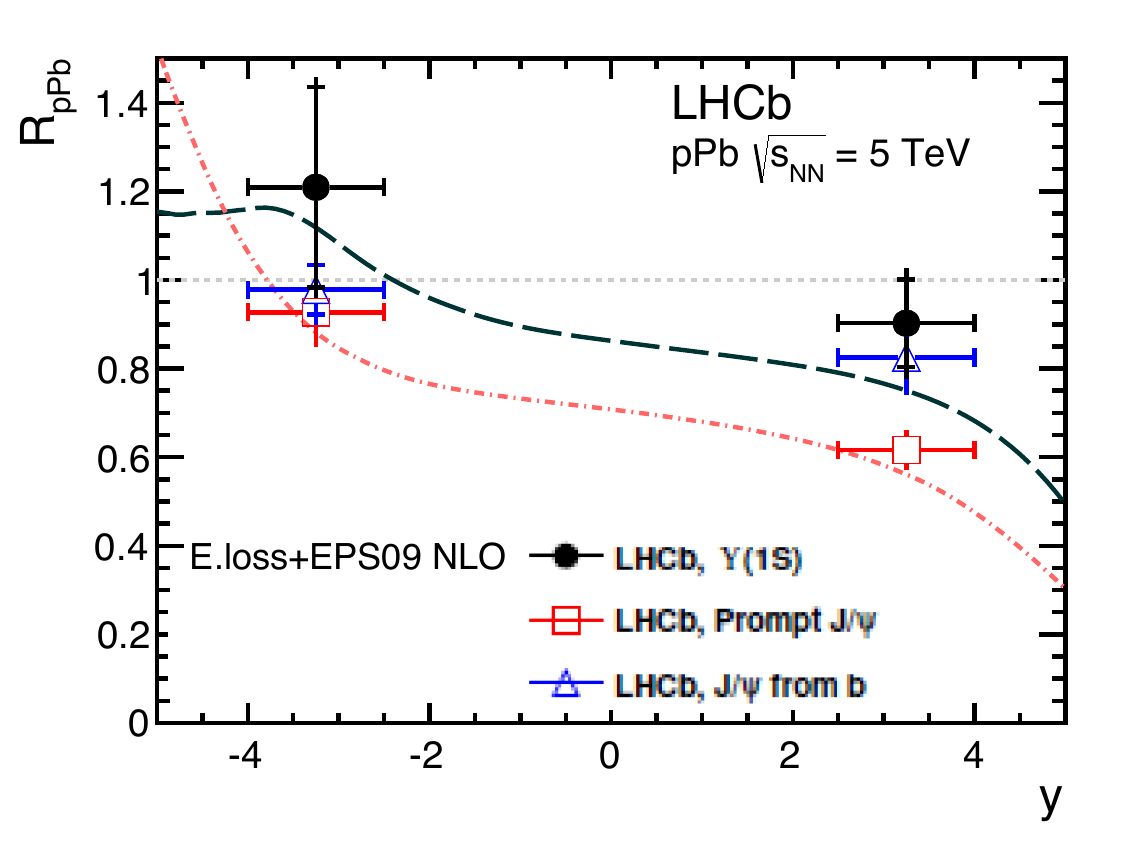}
 \includegraphics[width=5.0cm,clip]{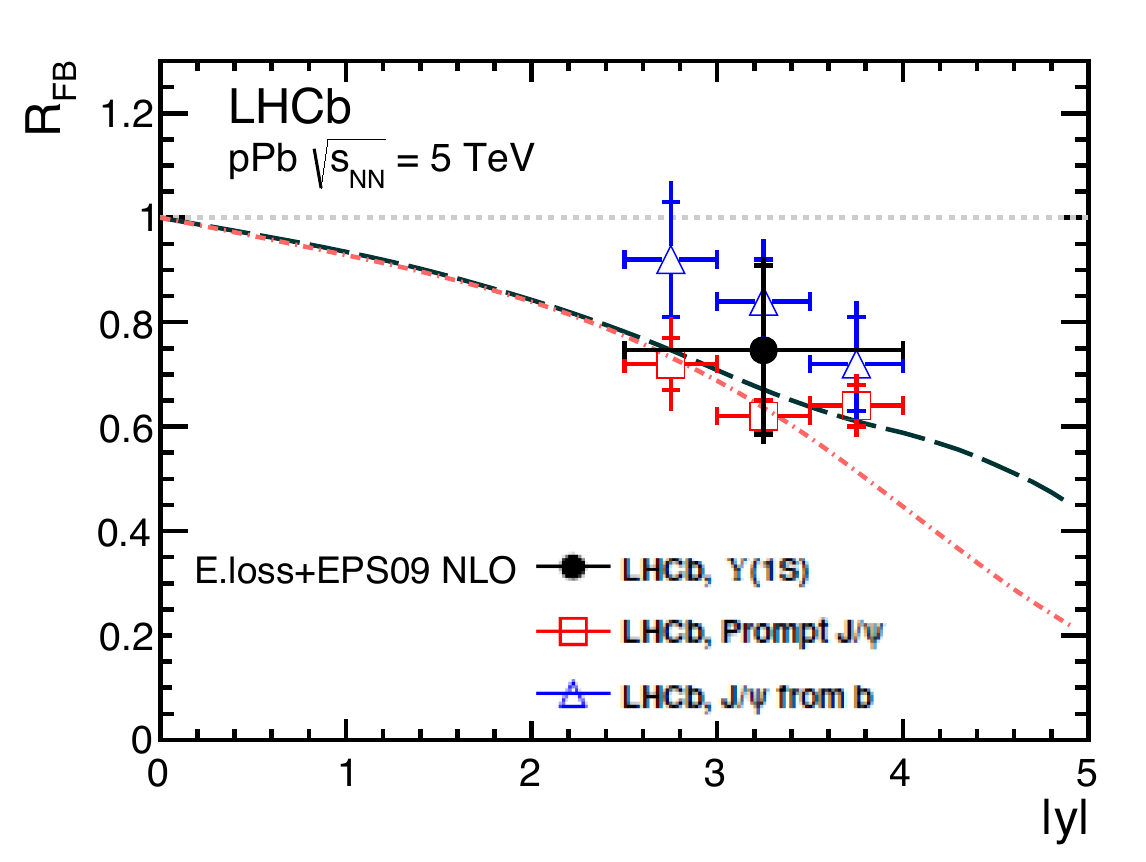}
\caption{$R_{\rm{pA}}$ (left) and $R_{\rm{FB}}$ (right) as a function of rapidity for prompt J/$\psi$ (red), J/$\psi$ from b (blue) and $\Upsilon$(1S) (black). The results are compared with theoretical predictions including energy loss and nuclear shadowing\cite{Arleo:2012rs}.}
\label{fig-3}       
\end{figure*}

Figure \ref{fig-2} left shows the $R_{\rm{pPb}}$ of prompt J/$\psi$ and $\psi$(2S) in the rapidity ranges -4.0 $< y <$ -2.5 and 2.5 $< y <$ 4.0 \cite{Aaij:2013zxa,LHCb:psi2015}. The results are compared to several theoretical calculations of parton shadowing and of coherent energy loss with or without shadowing\cite{Ferreiro:2013pua,delValle:2014wha,Albacete:2013ei,Arleo:2012rs}. A strong suppression of prompt J/$\psi$ is observed at the forward rapidity which is compatible with most of the theoretical predictions. The comparison of prompt J/$\psi$ and $\psi$(2S) $R_{\rm{pPb}}$ suggests that $\psi$(2S) are more suppressed than J/$\psi$, especially in the backward region. Models describing the J/$\psi$ data might still be able to describe the forward $\psi$(2S) suppression  but are not able to reproduce the suppression in the backward region. This intriguing result could be a first indication that another CNM effect is at play. Recently, a theoretical calculation based on comovers scenario \cite{Ferreiro:2014bia} tries to explain this behaviour. Figure \ref{fig-2} right shows the $R_{\rm{pPb}}$ of J/$\psi$ and $\psi$(2S) from b-hadrons in the rapidity ranges -4.0 $< y <$ -2.5 and 2.5 $< y <$ 4.0. J/$\psi$ from b are slightly suppressed in the forward region and both models including the shadowing effect can describe the data. This is the first indication of the b-hadron production in pPb collisions. In the backward region, J/$\psi$ from b are slightly less suppressed than prompt J/$\psi$ as expected from the models, however models are in worse agreement with the J/$\psi$ from b data. Given the large experimental uncertainties on the $\psi$(2S) from b measurement, no conclusions can be made on the comparison of the $\psi$(2S) from b suppression with respect to the J/$\psi$ from b suppression. Figure \ref{fig-3} left shows the measurement of $R_{\rm{pPb}}$ for $\Upsilon$(1S) as a function of rapidity \cite{Aaij:2014mza} compared with the $R_{\rm{pPb}}$ measurement of prompt J/$\psi$ and J/$\psi$ from b. $\Upsilon$(1S) is suppressed in the forward region while there is an indication for an enhancement of $\Upsilon$(1S) production with respect to pp in the backward region, which could be attributed to anti-shadowing. The $R_{\rm{pPb}}$ measurement of $\Upsilon(1S)$ agrees within uncertainties (albeit large in the backward region) with the $R_{\rm{pPb}}$ measurement of J/$\psi$ from b, reflecting the fact that CNM effects affect the b-hadrons production. $\Upsilon$(1S) data agree with coherent energy loss model including nuclear shadowing as parametrized with EPS09 \cite{Arleo:2012rs}. A comparison with more models is available in Ref. \cite{Aaij:2014mza}. Figure \ref{fig-3} right shows the forward-backward production ratio $R_{\rm{FB}}$ as a function of rapidity for $\Upsilon$(1S), prompt J/$\psi$ and J/$\psi$ from b. A smaller forward-backward asymmetry is observed for J/$\psi$ from b with respect to prompt J/$\psi$. The forward backward asymmetry measurement of $\Upsilon$(1S) and prompt J/$\psi$ agree with theoretical calculation of coherent energy loss with nuclear shadowing parametrized with EPS09\cite{Arleo:2012rs}.

\section{Z production in pPb and Pbp}

Z boson candidates have been reconstructed in the dimuon decay channel, in the fiducial region defined by 2.0 < $\eta(\mu^{\pm})$ < 4.5, $p_{T}(\mu^{\pm})$ > 20 GeV/c and 60 $< m_{\mu^{+}\mu^{-}} <$  120 GeV/$c^{2}$. A clean signal of 11 forward candidates in the forward region and 4 candidates in the backward region has been observed\cite{Aaij:2014pvu}. The cross section for Z production is found to be $\sigma_{Z \rightarrow \mu^{+}\mu^{-}} = 13.5^{+5.4}_{-4.0} (\rm{stat.}) \pm 1.2 (\rm{syst.})$~nb in the forward direction and $\sigma_{Z \rightarrow \mu^{+}\mu^{-}} = 10.7^{+8.4}_{-5.1} (\rm{stat.}) \pm 1.0 (\rm{syst.})$~nb in the backward direction. The result in the forward region agrees with predictions from NNLO calculations using FEWZ \cite{Gavin:2010az} and the MSTW08 PDF set \cite{Martin:2009iq} with and without nuclear effects. In the backward region, data are higher than theoretical calculations. The forward-backward ratio has also been measured in the rapidity range 2.5 $< \mid y_{CMS} \mid <$ 4.0. It is found to be lower than expectations with a 2.2$\sigma$ deviation from one.
The present measurements have a limited statistical precision, preventing to conclude on the presence of nuclear effects.

\section{Multi-particle correlations in pPb and Pbp}

\begin{figure*}
\centering
 \includegraphics[width=5.0cm,clip]{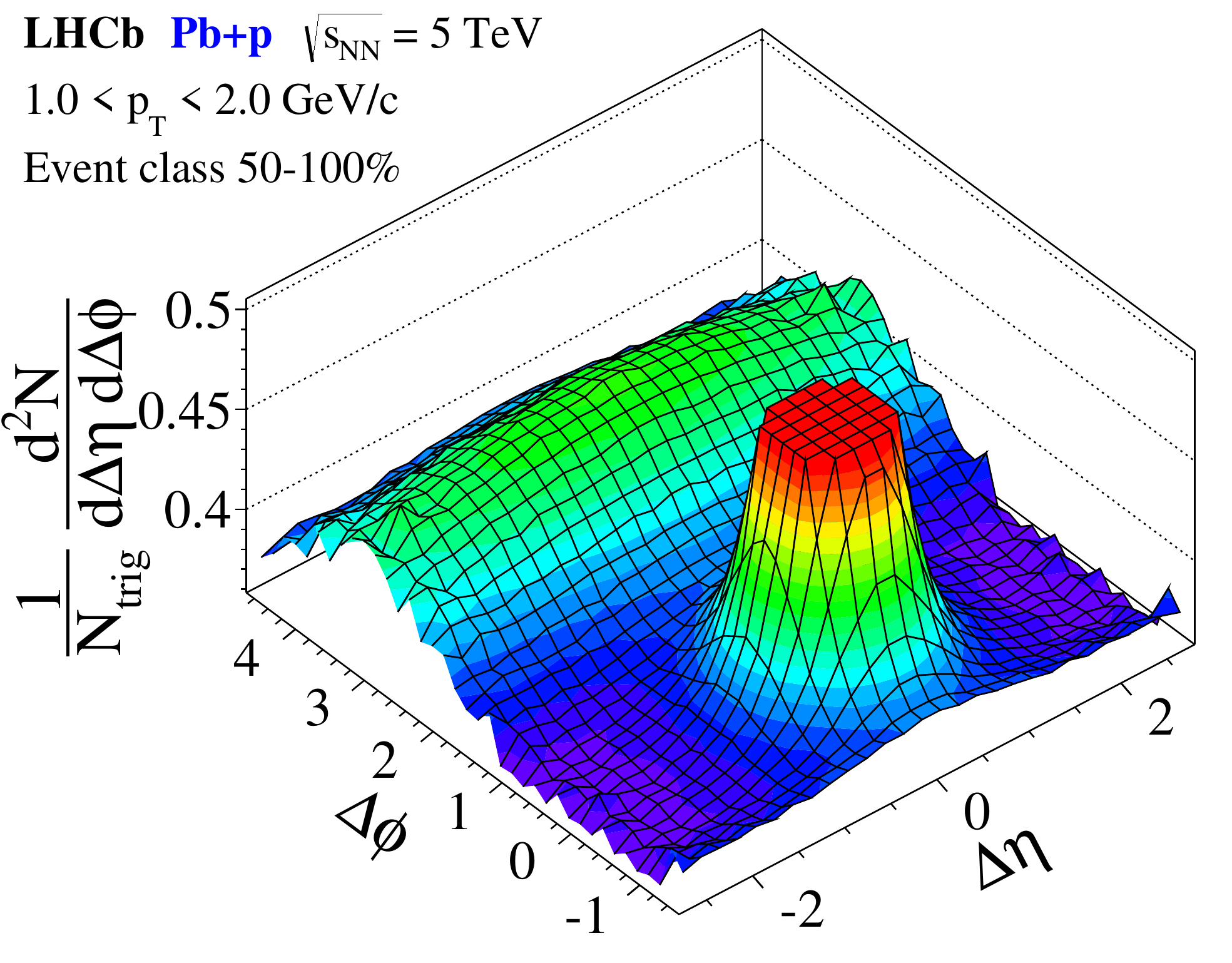}
 \includegraphics[width=5.0cm,clip]{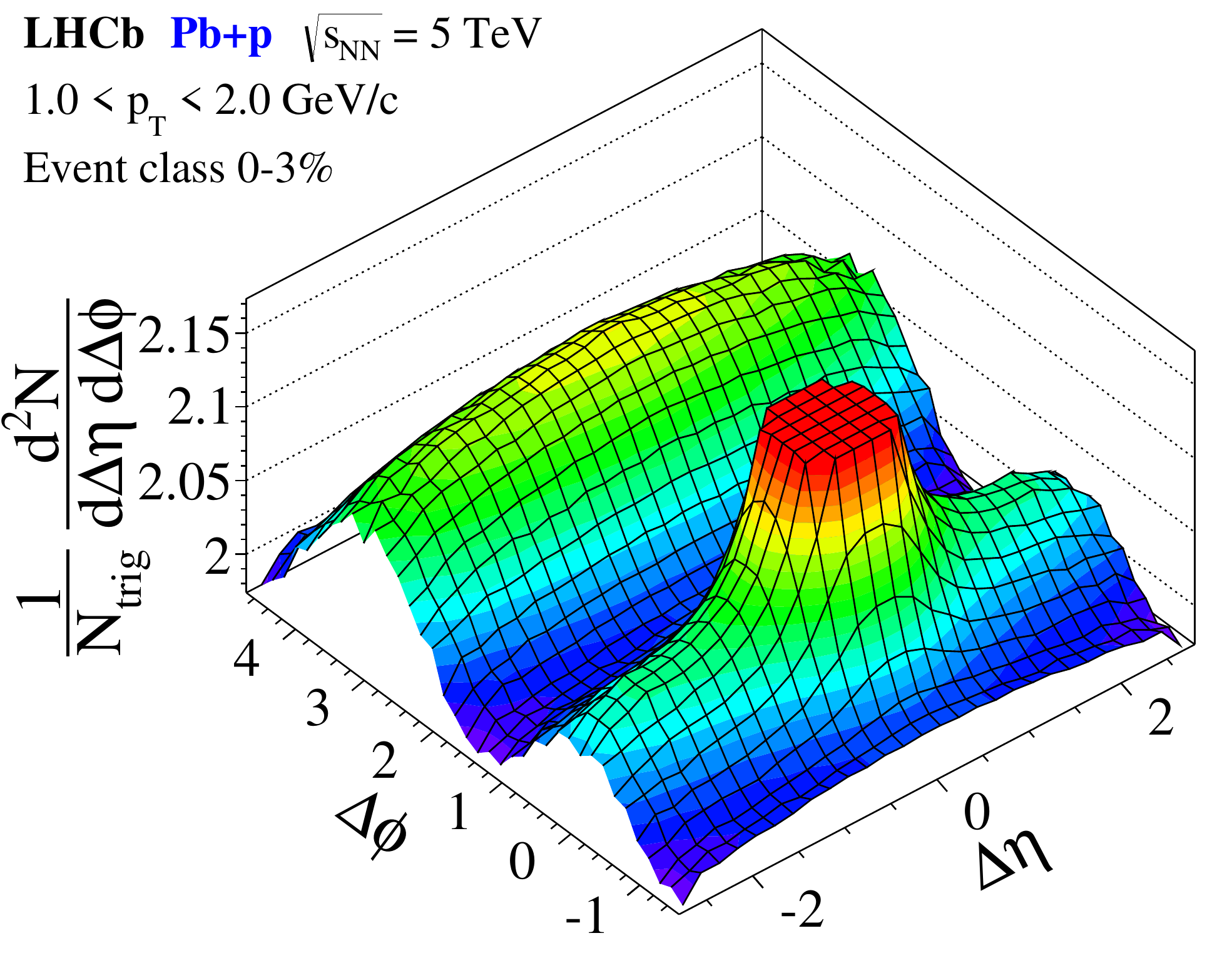}
\caption{Two-particle correlation functions in the backward Pbp configuration, for low event-activity class (left) and high event-acticity class (right). The analysed pairs of prompt charged particles are selected in the $p_{\rm T}$ range 1 < $p_{\rm T}$ < 2 GeV/$c$.}
\label{fig-4}       
\end{figure*}

Two-particle angular correlations between prompt charged particles \cite{Aaij:2015qcq} in pPb collisions have been measured in the forward (backward) region in the rapidity ranges 1.5 $< y <$ 4.4 (2.5 $< -y <$ 5.4), respectively. They were studied in event-activity classes, which are defined as fractions of the VELO hit-multiplicity distributions in a minimum bias sample. In events with high event-activity a long-range correlation on the near side, the so-called ridge, is observed both in the forward and backward configurations, for 1 $< p_{\rm T} <$ 2 GeV/$c$, while it is not observed in the 50-100$\%$ event-activity class (see Fig. \ref{fig-4} for the backward case). It was also observed that when probing identical absolute activity ranges in the pPb and Pbp configurations, the observed long-range correlations are compatible with each other (albeit there is a shift of one unit in rapidity coverage between both configurations).

\section{Summary and prospects for heavy ion studies}

LHCb sucessfully participated in the pPb data taking in 2013. The production of prompt J/$\psi$, $\psi(2S)$, J/$\psi$ and $\psi$(2S) from b and $\Upsilon$ has been measured in pPb collisions at $\sqrt{s_{NN}}$ = 5 TeV. J/$\psi$ from b and $\Upsilon$(1S) are less affected by cold nuclear matter effects than prompt J/$\psi$ which exhibits a stronger suppression. Prompt $\psi$(2S) seems even more suppressed than prompt J/$\psi$, especially in the backward region. Models of coherent energy loss with and without nuclear shadowing effects give a good description of the data, except for the production of $\psi$(2S) in the backward region. The prompt $\psi$(2S) data suggest that another CNM effect could be at play in the backward region. LHCb also performed the first measurement of Z production in pPb collisions at the LHC and the current analysis with limited statistical precision, will highly benefit from the larger statistics data sample which will be collected during run 2. Thanks to two-particle correlation analysis, LHCb observed for the first time the ridge in the forward region, in pPb collisions, thus complementing measurement at mid-rapidity from other experiments. The LHCb dectector will also collect for the first time PbPb data at $\sqrt{s_{NN}}$~=~5~TeV at the end of 2015. LHCb will thus lead a rich physics program covering heavy flavour physics, electroweak physics, soft QCD physics and Quark Gluon Plasma studies. It is expected to collect $L_{int} \approx$ 50-80 $\mu$b$^{—1}$ at the end of 2015. In addition, LHCb is in the unique position at the LHC to perform fixed target physics. Thanks to its System for Measuring Overlap with Gas (SMOG), initially used as luminosity monitor, noble gases can be injected inside the inner tracker of LHCb. With the proton and lead beams delivered by the LHC, and the various species of gas which can be injected in LHCb, a large number of fixed target configurations can be studied. LHCb already successfully collected proton-helium, proton-neon, proton-argon, lead-neon data and a lead-argon data taking is foreseen in parallel of the 2015 PbPb data taking. The highest center-of-mass energy which can be achieved in proton-gas collisions is $\sqrt{s_{NN}}$~=~115 GeV while in lead-gas collision it is $\sqrt{s_{NN}}$~=~72 GeV. Through its fixed target program, LHCb can bridge the gap from SPS to LHC with a single experiment. LHCb is definitely becoming a truly general purpose detector in the forward region.


\pdfoutput=1


%
%

\bibliography{templateISMD}

%
%
%

\end{document}